\documentclass[journal=jacsat,manuscript=article]{achemso}



\usepackage{chemformula} % Formula subscripts using \ch{}
\usepackage[T1]{fontenc} % Use modern font encodings

\author{Kazuki Morita}
\email{morita0@sas.upenn.edu}
\affiliation{Department of Chemistry, University of Pennsylvania, Philadelphia, Pennsylvania 19104-6323, USA}
\author{Andrew M. Rappe}
\email{rappe@sas.upenn.edu}
\affiliation{Department of Chemistry, University of Pennsylvania, Philadelphia, Pennsylvania 19104-6323, USA}

\title[]
  {
          How unconventional oxidation state Au$^{2+}$ is stabilized in halide perovskite \ch{Cs4Au3Cl12}:
          a first-principles study of its polaron crystal nature
  }

\begin{document}

\begin{abstract}
        Gold in crystalline compounds is typically only stable in oxidation states \ch{Au^{1+}} and \ch{Au^{3+}}.
        Even compounds with nominal \ch{Au^{2+}} usually disproportionate into \ch{Au^{1+}} and \ch{Au^{3+}}.
        Recently, \ch{Cs4Au3Cl12} was synthesized, where gold took the 2+ state in the bulk.
        Here, we investigate this compound using first-principles calculations and show that stabilization of the \ch{Au^{2+}} ion is through the formation of a polaron crystal.
        The electronic and phononic structure suggest that the bonding network can be interpreted as a collection of \ch{[Au^{2+}Cl4]^{2-}} and \ch{[Au^{3+}Cl4]^{1-}} square planar motifs, and the crystal lacks a smooth pathway for \ch{Au^{2+}} to disproportionate into \ch{Au^{1+}} and \ch{Au^{3+}} without creating dangling bonds.
        The electronic states of Au are contained within each \ch{AuCl4} motif, which allows for the \ch{Au^{2+}} state to be localized and isolated electronically.
        The \ch{Au^{2+}}-sites form an ordered structure, which is driven by a strong repulsive interaction between \ch{[Au^{2+}Cl4]^{2-}} motifs due to their lattice distortion.
        By considering redox reaction, we show that \ch{Cs4Au3Cl12} has the maximal density of \ch{Au^{2+}}, and further reduction will induce a delocalized state.
        \ch{Cs4Au3Cl12} has distinctive electronic structure, with a narrow gap, isolated HOMO and LUMO bands strongly localized at the Au-sites, and magnetization at the \ch{Au^{2+}}-sites making \ch{Cs4Au3Cl12} unique among quantum materials.
        \ch{Cs4Au3Cl12} can be a testbed to explore novel gold chemistry, opening opportunities to control oxidation state through engineering of lattice distortions.
\end{abstract}

\section{Introduction}

% XXX para1
% what different oxidation state can do
As one of the foundational concepts in chemistry, oxidation state plays a central role in scientific research since the early days.\cite{Pauling1960}
Indeed, ground-breaking materials development has often involved the realization of a particular oxidation state.\cite{Kageyama2018_NC,Walsh2018_NM}
Examples include enhancing the surface reactivity of catalysts,\cite{BaiY2022,DeP2018,JiangY2023,YangPP2020}
systematically tuning work functions,\cite{GreinerMT2012}
quantum criticality,\cite{MatsumotoY2011}
% and the emergence of superconductivity.\cite{WangY2014}\bibnote{Note that oxidation states can be ill-defined for metals.}
% \renewcommand{\thefootnote}{\roman{footnote}}
\renewcommand{\thefootnote}{\dag}
and the emergence of superconductivity.\cite{WangY2014}$^,$\footnote{Note that oxidation states can be ill-defined for metals.}
Even so, each atomic species has a limited number of stable oxidation states, and realizing any unusual oxidation state is still a challenge.

% XXX para2: about mixed-valence
Mixed-valence compounds are a particularly interesting class of materials, wherein a single constituent species takes multiple oxidation states.\cite{VarmaCM1976,DayP2007,KrickAL2016}
Unlike defects or dangling bonds at surfaces and interfaces, these species are homogeneously distributed across the material down to the unit cell level, and the disproportionated oxidation states are stable well beyond the time scale of phonon frequencies.\cite{VarmaCM1976}
% An intuitive and useful classification of mixed-valence compounds was proposed by Robin and Day, where class 1 is mixed-valence realized on different crystallographic sites, while class 3 has mixed valence on similar sites.\cite{DayP2007}
An intuitive and useful classification of mixed-valence compounds was proposed by Robin and Day, where class 1 is mixed-valence realized on different crystallographic sites, while class 2 has mixed valence on similar sites.\cite{DayP2007,VarmaCM1976}
The extreme of class 2 is class 3, where the sites become indistinguishable and the compound is no longer mixed-valence.
One clear example of class 1 is in $P2_1/c$ \ch{Sn3O4}, where \ch{Sn^{2+}} and \ch{Sn^{4+}} occupy the peak of a triangular pyramid and the center of an octahedron, respectively.\cite{ManikandanM2014}
On the other hand, an example of class 2 close to 3 is the spin polaron crystal (Zener polaron) seen in \ch{Pr_{0.6}Ca_{0.4}MnO3}, where lattice distortion and charge couple strongly to form a periodic structure beyond the periodicity of the original lattice.\cite{Daoud-AladineA2002}
Polaron crystals are an interesting example where, individual polarons gather to form a crystal, similar to the way molecules condense to form a molecular solid.
In their work, the environmental difference between \ch{Mn^{3+}} and \ch{Mn^{4+}} is attributed to polaronic lattice distortions.
Numerous other strategies exist to stabilize mixed-valence, such as doping,\cite{ZhouF2006} designing a specific metal-organic framework structure,\cite{ParkJG2018} or as exotic as using light to switch between different charge orderings.\cite{MatsumotoT2014}
% Since class 3 can involve a nominal non-integer oxidation state, the charges can be ordered or fully delocalized.
At present, the aim for mixed-valence materials is to realize class 1 or 2 mixed-valence materials and avoid class 3.
However, thermal fluctuations, lattice strain, redox reaction, and instability of the Fermi level can drive the system towards delocalized class 3.\cite{BrunschwigBS2002}
Therefore, to exhibit mixed-valence properties, compounds must be tolerant against these perturbations.

% XXX para3: Au specific
A recent noteworthy advance in the field of mixed-valence compounds is the synthesis of vacancy-ordered perovskite \ch{Cs4Au3Cl12} with \ch{Au^{2+}} and \ch{Au^{3+}}.\cite{Lindquist2023}
Based on M\"ossbauer spectroscopy, electron paramagnetic resonance, magnetic susceptibility measurements, and density functional theory (DFT) calculations, they report the stable existence of \ch{Au^{2+}}.
The authors also report distinctive flat band structure and enhanced electronic conductivity, and they suggest the possibility of exotic transport phenomena.
Not only is this material mixed-valence, but it is a rare case of \ch{Au^{2+}} in a solid-state system, where only a handful of other examples are reported.\cite{ElderSH1997,HwangIC2002}
% It should be noted that \ch{Au^{2+}} has been reported for complexes and molecules, but the fact that \ch{Cs4Au3Cl12} is a solid and not even a molecular solid makes it worthwhile to study.
The nominal electronic structure of $5d^9$ in \ch{Au^{2+}} creates an unpaired spin and makes \ch{Cs4Au3Cl12} a rare case where magnetism originates from gold.\cite{Lindquist2023}
The conventional perovskite \ch{CsAuCl3}, on the other hand, is also mixed-valence, but has \ch{Au^{1+}} and \ch{Au^{3+}} states.\cite{Ushakov2011,Debbichi2018}
Since the oxidation state of transition metal species is strongly entangled with lattice distortion, one might expect strain and pressure to induce \ch{Au^{2+}}, but both theory and experiment have shown that \ch{CsAuCl3} turns metallic before \ch{Au^{2+}} emerges.\cite{Winkler2001,Morita2024}
Therefore, the preference of gold to take 1+ and 3+ oxidation states is strong, and a strong mechanistic reason must be present in \ch{Cs4Au3Cl12} in order to stabilize the \ch{Au^{2+}} and prevent it from disproportionating.

% XXX para4: overview of the paper
In this work, we have investigated \ch{Cs4Au3Cl12} using first-principles calculations and show the validity of \ch{Au^{2+}} and mechanism that stabilizes it.
The structure of \ch{Cs4Au3Cl12} has gold-vacancy ordering that allows every gold to bond with four chlorine atoms and form a square-planar \ch{AuCl4} motif.
This bonding network lacks a smooth pathway to form \ch{Au^{1+}} without forming dangling bonds, providing a barrier against \ch{Au^{2+}} disproportionating into \ch{Au^{1+}} and \ch{Au^{3+}}.
The \ch{Au^{2+}} is identifiably different from \ch{Au^{1+}} and \ch{Au^{3+}} due to its unpaired spin.
The electronic and phononic band structures suggest that the \ch{Au^{2+}} states are largely confined within the \ch{[Au^{2+}Cl4]^{2-}} motif.
This independence allows \ch{Au^{2+}} to be stable as a form of isolated small polaron state, where interaction with the other \ch{Au^{2+}} ions is minimal.
At the same time, however, the \ch{[Au^{2+}Cl4]^{2-}} motif, as a whole has a repulsive strain-field interaction to the other \ch{[Au^{2+}Cl4]^{2-}}, resulting in the \ch{Au^{2+}} ions exhibiting an ordered structure.
Unlike many mixed-valence compounds, we also show that \ch{Cs4Au3Cl12} is stable against thermal fluctuations and redox reactions.
Our results establish the validity and stability of \ch{Au^{2+}} in \ch{Cs4Au3Cl12} and that the stabilization mechanism is explained by the formation of a polaron crystal.

%-------------------------------------------------------------------------------
\section{Results and Discussion}

%------------------------------------------------------------------------------
% structure
%\subsection{Structure and bonding}

\begin{figure}
  \includegraphics[width=130mm]{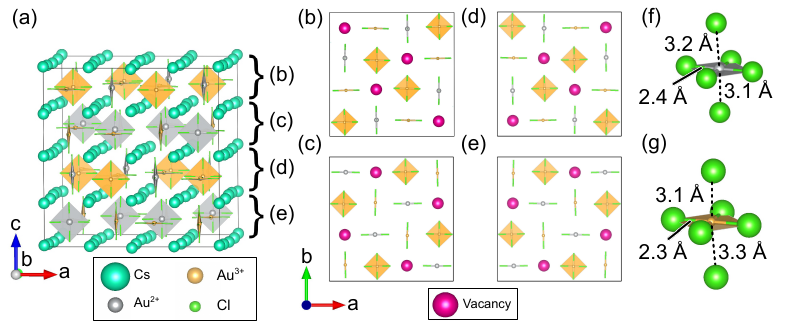}
        \caption{
                (a) Conventional unit cell of \ch{Cs4Au3Cl12}.
                (b)$\sim$(e) Cross section along successive (001) planes.
                (f) Distances to six nearest Cl around (f) \ch{Au^{2+}} and (g) \ch{Au^{3+}}.
        }
  \label{fig:struct} % fig1
\end{figure}

{\bf Structure and bonding.} The DFT optimized structure of \ch{Cs4Au3Cl12} is shown in Figure~\ref{fig:struct}(a), where it takes the tetragonal $I4_1cd$ (\#110) space group.
The structure can be understood as a vacancy-ordered perovskite, where one out of every four Au atoms is missing from \ch{CsAuCl3}, forming $4\times4\times4$ superlattice.
The Au-site is mixed-valence, taking oxidation states of \ch{Au^{2+}} and \ch{Au^{3+}}, which are presented in grey and yellow in Figure~\ref{fig:struct}, respectively.
Both Au-sites have six neighboring Cl-sites, but two of the Cl-sites are far ($\sim$35\% farther) and beyond the Cs-plane, therefore making it effectively a square-planar \ch{AuCl4} bonding environment (Figures~\ref{fig:struct}(f) and (g)).
We, therefore, depict Au-sites as square \ch{AuCl4} motifs showing the bonding to the four closest Cl-sites for the rest of the work.
The cross-sectional views of the conventional cell along (001) planes are shown in Figure~\ref{fig:struct}(b)-(e).
One can find that any linear $4\times1$ set of Au-sites taken along any of the axes has two \ch{Au^{3+}}-sites, a \ch{Au^{2+}}-site, and a vacancy.
None of the square-planar \ch{[Au^{2+}Cl4]^{2-}} motifs lie in the (001); they lie in the (100) plane (Figure~\ref{fig:struct}(b) and (d)) or the (010) plane (Figure~\ref{fig:struct}(c) and (e)).
% In each plane, the \ch{Au^{3+}Cl4} has two crystallographic orientations.
% In each plane, half the \ch{Au^{3+}Cl4} are perpendicular to both the (001) plane and the \ch{Au^{2+}Cl4}, and the other lies in the (001) plane.
%Furthermore, any \ch{[AuCl4]} has two vacancy cells next to it along its motif normal directions.
Furthermore, any \ch{[AuCl4]} has two vacancy cells next to it along two of its in-plane directions.
At first glance, the crystal structure seems complicated, as though a simpler crystal structure could exist.
However, we show that the structure is uniquely determined by two conditions.
% Firstly, if we ignore the difference between the \ch{Au^{2+}} and the \ch{Au^{3+}} and only consider the connectivity of Au and Cl-sites, the structure is uniquely determined by packing the \ch{AuCl4} motifs and the Au-vacancies into the Cs lattice while making sure that all \ch{AuCl4} has two Au-vacancies next to it.
% In other words, this bonding network of \ch{Cs4Au3Cl12} is the only one that can host the \ch{AuCl4} motifs.
Firstly, if we ignore the difference between the \ch{Au^{2+}} and the \ch{Au^{3+}} and only consider the connectivity of Au and Cl-ions, the structure is uniquely determined by assuring that each \ch{AuCl4} has two Au-vacancies next to it (step-by-step explanation presented in Supplementary Information).
Secondly, the \ch{Au^{2+}}-site locations are determined by making sure that they do not neighbor each other.
There are three unique ways to choose this, and each of the choices corresponds to selecting one axis as the $c$-axis for the tetragonal $I4_1cd$ structure.
The tendency for \ch{Au^{2+}} to repel each other was confirmed by calculating an artificial structure with two neighboring \ch{Au^{2+}}-sites and finding it to be unstable. % TODO add result to SI
% However, we suggest this repulsion to be short-ranged and is not strong enough to destabilize the crystal itself.
These two rules uniquely determine the \ch{Cs4Au3Cl12} structure.

Here, the Au-vacancy plays a dual role of making enough Cl-sites available for gold to form the \ch{AuCl4} motif and absorbing the strain caused by the larger size of \ch{Au^{2+}}.
In particular the two Au-vacancies next to each \ch{[Au^{2+}Cl4]^{2-}} motif allow the \ch{Au^{2+}}-Cl bonds to relax along the in-plane directions and prevent the strain from propagating to the neighboring \ch{AuCl4} motifs.
Every Cl ion belongs to a \ch{AuCl4} motif, so a smooth pathway to form \ch{Au^{1+}}, where it takes linear bonding environment,\cite{Morita2024} is not possible without forming an energetically unfavorable isolated Cl.
The above geometrical argument explains the structural stability of \ch{Cs4Au3Cl12}, but it is based on a rudimentary ball-and-stick model.
The validity of this simplified picture will be justified in the following sections.

%------------------------------------------------------------------------------
% electronic structure
% \subsection{Electronic structure}

\begin{figure}
  \includegraphics[width=110mm]{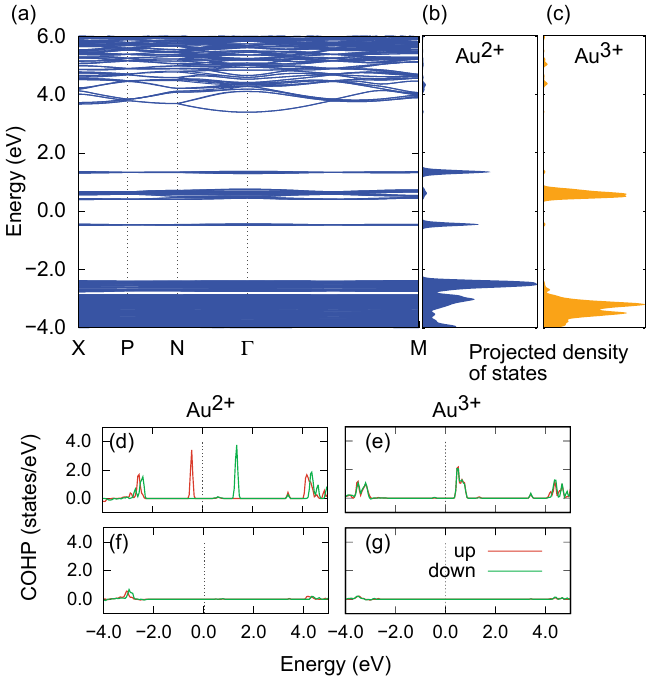}
        \caption{
                (a) Electronic band dispersion. Zero of energy is taken to be the center of the band gap.
                Projected density of states for (b) \ch{Au^{2+}} and (c) \ch{Au^{3+}}.
                Spin-resolved crystal orbital Hamilton population (COHP) of
                (d) nearest-neighbor \ch{Au^{2+}-Cl},
                (e) nearest-neighbor \ch{Au^{3+}-Cl},
                (f) second nearest-neighbor \ch{Au^{2+}-Cl}, and
                (g) second nearest-neighbor \ch{Au^{3+}-Cl}.
                Spin up and down are plotted with red and green, respectively.
        }
  \label{fig:elec} % fig2
\end{figure}

% isolated flat bands
% COHP
{\bf Electronic structure.} \ch{Cs4Au3Cl12} exhibits a distinctive electronic structure, where the valence and the conduction band edges are isolated from the rest of the bands (Figure~\ref{fig:elec}(a)).
The isolated valence band edge is mainly composed of the \ch{Au^{2+}} (Figure~\ref{fig:elec}(b)), and the lowest conduction band is mainly composed of the \ch{Au^{3+}} (Figure~\ref{fig:elec}(c)) with the band gap being 0.85 eV.
The density of states was reported previously by Lindquist et al., and our result closely agrees with theirs.\cite{Lindquist2023}
These states exhibits a small dispersion, suggesting that the gold states are highly localized.
The band splitting in \ch{Au^{2+}} is the result of the ${\rm d}_{x^2-y^2}$ orbital becoming half-occupied.
To elucidate the contribution of these states towards bonding, the parity of the wavefunction was analyzed using spin-resolved crystal orbital Hamilton population (COHP), which takes a positive value when the interaction between atoms is bonding.\cite{Maintz2016}
Both \ch{Au^{2+}} and \ch{Au^{3+}} were only bonded to the nearest Cl-sites, and little interaction was present between Au and the second-nearest-neighbor Cl-sites. (Figure~\ref{fig:elec}(d)-(g)).
This supports the geometrical model that the crystal structure of \ch{Cs4Au3Cl12} can be discussed based on the \ch{AuCl4} motif.

% some properties
% oxidation state: Bader Mulliken
% Born effective charge
% XXX little magnetic interaction, backed by experimental response (?)
% spin
% energy difference (?), experimental value?
% JiangL2012 analysis
We next assessed the validity of the \ch{Au^{2+}} state by analyzing the charge around the \ch{Au^{2+}}-site.
The charge assignment was done via the wavefunction topology and the modern theory of polarization method.\cite{JiangL2012}
We created a series of structures where four \ch{Au^{2+}} ions are gradually displaced to the next \ch{Au^{2+}}-site along the z-direction, equivalent to one \ch{Au^{2+}} being displaced along the full length of the calculation cell in the z-direction (detailed path in supporting information).
The current caused by this modification was evaluated using the Berry phase calculation, and the charge carried by \ch{Au^{2+}} was indeed 2+.
We also found that local positive charge was 30\% larger around \ch{Au^{3+}} than that around \ch{Au^{2+}} using Bader charge analysis.
Interestingly, the Born effective charges are very anisotropic, and the out-of-plane component was small (Table~S3).
More explicitly, \ch{Au^{2+}} exhibits a finite magnetic moment of 0.43 $\mu_{B}$ (\ch{Au^{1+}} and \ch{Au^{3+}} show $\mu_B=0$), which was also reported in the previous work.\cite{Lindquist2023}
This is a qualitative and quantitative difference only found for \ch{Au^{2+}}.
We further investigated the magnetic interaction between \ch{Au^{2+}} ions.
Six different antiferromagnetic and ferrimagnetic configurations differed less than 5.5 meV in energy (Table~S4), suggesting that the magnetic interaction between \ch{Au^{2+}} ions is weak and that a large number of different magnetic configurations are likely to be accessible even in low temperatures.

% polarisation, symmetry discussion
\ch{Cs4Au3Cl12} belongs to a polar $I4_1cd$ space group and has a finite spontaneous polarization of 0.056 C/m$^2$.
Interestingly, the spontaneous polarization is solely attributed to the difference between \ch{Au^{2+}} and \ch{Au^{3+}}-sites, therefore almost purely electronic in nature.
The spontaneous polarization can be reversed by \ch{Au^{2+}} and \ch{Au^{3+}}-site exchanging their locations.
% This is close to the electronic polarization concept suggested in some previous works.\cite{Morita2022_PRL}
These properties also augment the quantitative difference between \ch{Au^{2+}} and \ch{Au^{3+}}.
From the above results, we believe it is legitimate to consider \ch{Au^{2+}} as 2+ oxidation state.
Lindquist et al. suggested \ch{Cs4Au3Cl12} to be classified as class 2 in the Robin and Day classification.\cite{Lindquist2023}
Our results also show the qualitative distinctiveness of \ch{Au^{2+}} and \ch{Au^{3+}}.
At the same time, the local environments around the \ch{Au^{2+}}-site and the \ch{Au^{3+}}-site are similar, and thus, \ch{Cs4Au3Cl12} is class 2, close to class 3.

%------------------------------------------------------------------------------
% phonon structure
% \subsection{Phononic structure}

\begin{figure}
  \includegraphics[width=65mm]{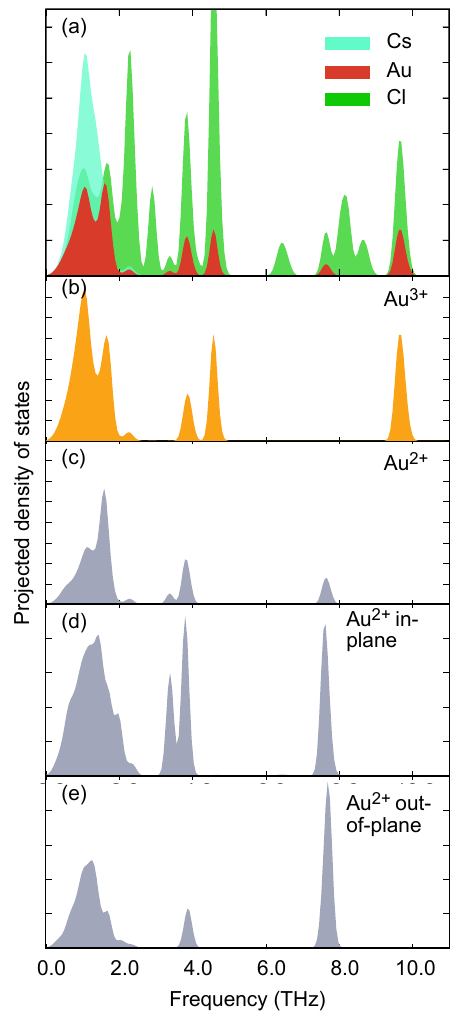}
        \caption{
                (a) Phonon-projected density of states for Cs, Au, and Cl.
                Phonon-projected density of states for (b) \ch{Au^{2+}} and (c) \ch{Au^{3+}}.
                Phonon-projected density of states for (d) in-plane and (e) out-of-plane \ch{Au^{2+}-Cl} direction.
        }
  \label{fig:phonon} % fig3
\end{figure}

% XXX points to raise
% lack of imaginary mode
% flat phonon bands
% little Cs contribution
% non-changing chemical property
% non-bonding (longer bond) is has higher frequency

{\bf Phononic structure.} Next, we look at the dynamical properties through the phonons.
The projected phonon density of states (Figure~\ref{fig:phonon}(a)) suggests that the Cs vibration is localized around 1.0 THz, and the bands above 2.0 THz are mostly attributed to the Au and Cl atoms.
This suggests that Cs is dynamically isolated from the rest of the lattice, as seen in perovskites.\cite{Morita2022_CM}
Cl peaks are distributed widely across the spectra, but breathing modes around Au and out-of-plane modes are observed without contributions from Cs or Au.
Au-Cl interactions are largely encoded in the three groups of peaks above 2.0 THz, which had simultaneous contributions from Au and Cl: three peaks around 4.0 THz, a peak around 8.0 THz, and a peak around 10.0 THz.
Within the Au spectra, \ch{Au^{2+}} had consistently lower frequencies compared to \ch{Au^{3+}} (Figures~\ref{fig:phonon}(b) and (c)).
This is explained by the suppressed interaction between the \ch{Au^{2+}} and Cl, due to the longer inter-atomic distance caused by the polaronic distortion.
Further decomposition of the spectrum into in-plane direction (Figure~\ref{fig:phonon}(d)) and out-of-plane direction (Figure~\ref{fig:phonon}(e)) revealed that the lowest frequency peak around 4.0 THz was solely attributed to \ch{Au^{2+}} vibrating in the in-plane directions.
This suggests significantly stronger interaction within the \ch{Au^{2+}Cl4} motif compared to the interaction with the neighbouring \ch{AuCl4} motif.
We also did not find any imaginary modes, suggesting \ch{Cs4Au3Cl12} to be stable against thermal perturbations at low temperatures (Figure~S10(a)).
\ch{Cs4Au3Cl12} is a soft material, similar to halide perovskites,\cite{ZachariasM2023} so we expect anharmonic phonons to play a non-negligible role at room temperature.
Therefore, the phase stability calculation at higher temperatures will require higher-order phonon calculations than this study.

\begin{figure}
  \includegraphics[width=130mm]{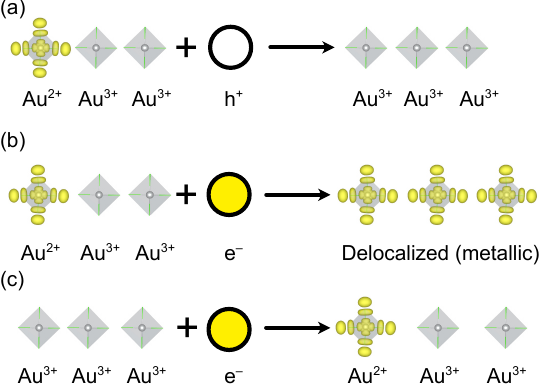}
        \caption{
                Schematic images of reactions involving \ch{Au^{2+}}.
                (a) oxidation,
                (b) reduction,
                and (c) reverse reaction of (a).
        }
  \label{fig:defect} % fig4
\end{figure}

% results
% stability
{\bf Redox stability.} Since many mixed-valence compounds are unstable against redox reactions, we have analysed the redox stability of \ch{Cs4Au3Cl12} by calculating the energy to add an electron or a hole to \ch{Au^{2+}}.
First, we consider a situation where \ch{Cs4Au3Cl12} is oxidised (Figure~\ref{fig:defect}(a)).
When an extra hole was introduced, it localized at the \ch{Au^{2+}}-site, forming \ch{Au^{3+}}.
% We found that the introduced extra hole localized at the \ch{Au^{2+}}-site, forming \ch{Au^{3+}}:
% \begin{equation}
% \label{rct:redox1}
% \ch{Cs4{Au^{2+}}{Au^{3+}2}Cl12 + h^+ -> Cs4{Au^{3+}}3Cl12.}.
% \end{equation}
% The reaction was an exothermic process with 0.40 eV energy difference per \ch{Au^{2+}} (which is per formula unit), suggesting it occurs spontaneously.
The reaction was an exothermic process with 0.40 eV, suggesting it occurs spontaneously.
This reaction also indicates the energy released by freeing the lattice distortion caused by an \ch{Au^{2+}} polaron, which corresponds to the polaron formation energy.
Next, we consider a reduction of \ch{Cs4Au3Cl12} (Figure~\ref{fig:defect}(b)), where we consider adding an electron into the conduction band.
Instead of forming an extra \ch{Au^{2+}}, the extra electron remained delocalized and formed a metallic state.
% \renewcommand{\thefootnote}{\ddag}
% Instead of forming an extra \ch{Au^{2+}}, the extra electron delocalized and formed a metallic \ch{Cs4{Au^{2.33+}}3Cl12},\footnote{Note that the oxidation state of \ch{Au^{2.33+}} is a nominal value.}
% \begin{equation}
% \label{rct:redox2}
%         \ch{Cs4{Au^{2+}}{Au^{3+}2}Cl12 + e^- -> Cs4{Au^{2.33+}}3Cl12\ (metallic).}
% \end{equation}
The reaction did not release or absorb any energy, and the electron simply remained in the conduction band.
% Instead of extra electron localizing to form \ch{Cs4{Au^{2+}}2{Au^{3+}}Cl12}, the extra electron delocalized and formed a metallic \ch{Cs4Au3Cl12}.
This suggests that the \ch{Au^{2+}} density in \ch{Cs4Au3Cl12} is saturated and that additional attempts to add an electron will cause the occupation of the conduction band.
% Finally, we also consider the inverse reaction of Equation~\ref{rct:redox1} and consider adding a conduction band electron to \ch{Cs4{Au^{3+}}3Cl12} (Figure~\ref{fig:defect}(c)).
Finally, we also consider the inverse reaction of Figure~\ref{fig:defect}(a).
Here we prepared a system where one \ch{Au^{2+}} is already oxidized to \ch{Au^{3+}} (Figure~\ref{fig:defect}(c)).
Note that this differs from Figure~\ref{fig:defect}(b) because the reaction starts with the oxidized state.
In contrary to Figure~\ref{fig:defect}(b), the extra electron localized at \ch{Au^{3+}}-site forming \ch{Au^{2+}}.
% Note that this differs from Equation~\ref{rct:redox2} because the reaction starts with the oxidized state.
% In contrary to Equation~\ref{rct:redox2}, the extra electron localized at \ch{Au^{3+}}-site forming \ch{Au^{2+}}:
% \begin{equation}
% \label{rct:redox3}
%         \ch{Cs4{Au^{3+}}3Cl12 + e^- -> Cs4{Au^{2+}}{Au^{3+}}2Cl12.}
% \end{equation}
This reaction is exothermic with 0.48 eV energy difference per \ch{Au^{2+}}.
% This result suggest that when \ch{Cs4Au3Cl12} is oxidised due to defects or doping, an extra electron added to the system can occupy \ch{Au^{3+}} defect site to form \ch{Au^{2+}}.
This further supports that \ch{Au^{2+}} density in \ch{Cs4Au3Cl12} is saturated.

The above results highlight the largely asymmetric nature of the \ch{Au^{2+}} state stability against redox reactions.
Figure~\ref{fig:defect}(a) indicates that the removal of electrons can be done locally without destroying other \ch{Au^{2+}}, which suggests that \ch{Cs4Au3Cl12} is stable against modest oxidation.
On the other hand, Figure~\ref{fig:defect}(b) indicates that reduction does not create additional \ch{Au^{2+}}, but instead makes the crystal metallic.
This suggests vulnerability of \ch{Au^{2+}} against reduction.
However, Figure~\ref{fig:defect}(c) suggests that if the crystal is already oxidized, then some additional \ch{Au^{2+}} can be created.
Therefore, we suggest that \ch{Cs4Au3Cl12} should be synthesized to be slightly \ch{Au^{2+}}-deficient and Cl-excess chemical conditions (oxidized), such as \ch{Cs4Au^{2+}_{1-x}Au^{3+}_{2+x}Cl_{12+x}}, where $x$ is greater than zero, to retain the mixed-valence state.

Finally, it is worthwhile to summarize and revisit the nature of \ch{Cs4Au3Cl12} as a polaron crystal.
A polaron is a complex of a charge carrier and a lattice distortion, which can be considered as a single composite particle.
Just like molecules gathering to form a molecular crystal, a polaron crystal (or sometimes referred to as a polaron solids) is collection of individual polarons.\cite{Daoud-AladineA2002,JoossC2007,YamadaY1996}
It is this individualistic characteristic of polarons that distinguish them from other continuous materials.
% In \ch{Cs4Au3Cl12}, the electronic state of \ch{Au^{2+}} is fully contained within the \ch{[Au^{2+}Cl4]^{2-}} motif, but at the same time the surrounding lattice distortion creates a strong strain field that causes \ch{[Au^{2+}Cl4]^{2-}} motifs to repel each other.
\ch{Cs4Au3Cl12}, however, also has unique polaron behavior compared to polaron crystals in manganese oxides.\cite{MannellaN2007, JoossC2007,YamadaY1996,Daoud-AladineA2002}
The distinctive behavior of a polaron crystal was suggested to be its temperature dependence, where the polaron crystalline order can melt to change the observed transport properties drastically.\cite{MannellaN2007, JoossC2007}
In \ch{Cs4Au3Cl12}, however, such a polaron melting was not observed experimentally,\cite{Lindquist2023} suggesting the \ch{Au^{2+}} to be more stable than other polaron crystals.
The \ch{Au^{2+}} states in \ch{Cs4Au3Cl12} were largely independent electronically and magnetically; however, the strain field created by each \ch{[Au^{2+}Cl4]^{2-}} motif interacted strongly and resulted in the ordered \ch{Au^{2+}} structure.
This is in contrast to the case of manganse oxides, where strong electronic and magnetic interactions dominate the polaronic behavior\cite{YamadaY1996,Daoud-AladineA2002}.
The fact that the \ch{Au^{2+}} cannot exist without Au-vacancy suggests it to be more similar to a polaron trapped near a defect rather than a self-trapped polaron.
The approach of using lattice distortion to control oxidation state is likely to be valid for other ionic species.
Some examples include \ch{Ag^{2+}}, \ch{In^{2+}}, and \ch{Tl^{2+}}, where different oxidation states have been reported in double perovskite.\cite{ZhangL2019_AEL,Retuerto2013,ZhaoXG2018,Wolf2022,XiaoZ2017_JPCL}

% % EI
% A further interesting point is the possibility of this material being an excitonic insulator.
% Excitonic insulator is an ordered state of electron-hole pairs, where one of the their indicator is a larger binding energy than the band gap.\cite{KanekoT2025}
% Low-dimensional systems have been suggested to be favourable, and \ch{Cs4Au3Cl12} is effectively a zero-dimensional material.
% The electronic band structure of \ch{Cs4Au3Cl12} (Figure~\ref{fig:elec}(a)) suggests a strong binding energy between the electron-hole pair generated at \ch{Au^{2+}}-site.
% The band edges are isolated from the rest of the band, so the excitons will reside only within the Au-site.
% The conduction band edge is \ch{Au^{3+}}-site; however, in order for an electron to decay to \ch{Au^{3+}}-site, it has to distort the lattice and escape the Coulomb attraction from a hole at \ch{Au^{2+}}, which is suggested to be a slow process.
% Therefore, long-lived excitons can exist in this system by having both electrons and holes in the \ch{Au^{2+}} site, but it is difficult to recombine.

%-------------------------------------------------------------------------------
\section{Conclusion}

% Oxidation states are conceptually simple yet chemically complex.
We have shown that lattice distortion can be a useful lever to control oxidation state.
Structurally \ch{Cs4Au3Cl12} is a collection of independent \ch{Cs^+}, \ch{[Au^{2+}Cl4]^{2-}}, and \ch{[Au^{3+}Cl4]^{1-}} sites, where bonds between these entities are weak.
The structure lacks a smooth pathway toward \ch{Au^{1+}} formation, which prevents the \ch{Au^{2+}} from disproportionating.
We provided evidence that \ch{Au^{2+}} is indeed a 2+ through charge analyses and observation of a finite magnetic moment.
The localized \ch{Au^{2+}} state was stabilized through strong trapping of a hole at each adjacent \ch{Au^{3+}} site, suggesting this crystal to be interpreted as a polaron crystal.
The \ch{Au^{2+}} state is likely to be stable against thermal and electronic perturbation.
Each \ch{Au^{2+}} is electronically and magnetically independent, but at the same time, the strain-field interaction is strong, leading to repulsion between \ch{Au^{2+}} ions.
% Our calculations also suggested that \ch{Cs4Au3Cl12} has a saturated \ch{Au^{2+}} density, and the addition of more electrons is predicted cause delocalized conduction band states to be occupied.
Our calculations also suggested that \ch{Cs4Au3Cl12} has a saturated \ch{Au^{2+}} density, and reduction is predicted to cause it to become metallic.
% The perculiar band structure and strong electron phonon coupling makes \ch{Cs4Au3Cl12} a unique material that is likely to have astrong exciton coupling.
Furthermore, the rare Au magnetism in \ch{Cs4Au3Cl12} will open an avenue for gold magnetism chemistry, which is still largely unexplored.
Overall, this work supports the validity of \ch{Au^{2+}} suggested by Lindquist et al., and this work also suggests the polaron crystal as a good model to explain the behavior of \ch{Au^{2+}} in \ch{Cs4Au3Cl12}.
Coupling with lattice distortion is common across transition-metal and post-transition-metal elements and suggests that an analogous approach could lead to polaron crystal design for accessing unconventional oxidation states and novel chemistry in many other compounds.

%-------------------------------------------------------------------------------
\section{Methodology}

Pseudopotential plane-wave DFT calculations were performed using the VASP code.\cite{Kresse1996_PRB11169,Kresse1999}
For reciprocal lattice sampling, a Monkhorst-Pack sampling was adopted, with spacing at most 0.28~\AA$^{-1}$\ between the neighboring k-points and a plane-wave cutoff of at least 400 eV.
9, 11, and 7 valence and/or semi-core electrons were considered for Cs, Au, and Cl, respectively.
The hybrid exchange-correlation functional (HSE06) was used throughout the work,\cite{Heyd2003} except for the phonon calculation, where PBE+$U$ ($U$=3.4 eV) was used.
The Hubbard $U$ was applied to the Au 5d-orbital, and the rotational invariant approach suggested by Dudarev et al. was used.\cite{Dudarev1998,KulikHJ2015}
Phonon modes were calculated in a 152-atom primitive cell using the phonopy package, and eigenvectors of the phonon bands were projected onto an atomic basis to analyze their contributions.\cite{Togo2023_JPCM,Togo2023_JPSJ}

The initial structure for the DFT calculations was taken from a structure reported by Karunadasa and co-workers\cite{Lindquist2023} and was subsequently relaxed.
Spontaneous polarization was calculated using the Berry phase method in the modern theory of polarization.\cite{Zak1989,Resta1992,King-Smith1994}
An inverse-polarization pair was generated by creating a mirror image structure where an electron at the \ch{Au^{2+}} hops two site along $-z$ direction to a \ch{Au^{3+}}-site.
The narrow band gap of this material made the perturbative approach numerically unstable, so the Born effective charge was calculated by a finite displacement method.
Symmetry analysis was done using Spglib and ISOTROPY software.
The bader charge analysis was performed with analysis code developed by Henkelman and co-workers.\cite{HenkelmanG2006}
COHP was calculated by the LOBSTER package, where six Cl-sites around all Au-sites were considered.\cite{Maintz2016}

The wavefunction topology and the modern theory of polarization method was used to analyze the nominal charge of \ch{Au^{2+}}.\cite{JiangL2012}
We displaced four \ch{Au^{2+}} along the z-direction to the next \ch{Au^{2+}}-site so that each \ch{Au^{2+}} moves 25\% of the calculation cell (more details in supporting information).
In other words, a current was generated equivalent to moving \ch{Au^{2+}} by one full supercell lattice vector in the z-direction, restoring the original structure through periodic boundary conditions.
The initial and the final structures were equivalent, so the change in the Berry phase is always an integer multiple of $2\pi$, or $n$ times the polarization quantum.
This $n$ plus the nuclear charge is the oxidation state of the ion.

Redox reaction was modeled by calculating charged systems, where +1 and -1 was considered for oxidized and reduced \ch{Cs4Au3Cl12}, respectively.
As commonly done in periodic boundary conditions, we calculated charged systems with a neutralizing background charge.
In experiment, equivalent situation would be caused by a uniform potential such as by gating.
To remove spurious interactions between periodic image charges, we applied the Kumagai-Oba correction (extended Freysoldt-Neugebauer-Van de Walle correction).\cite{Freysoldt2009,Kumagai2014_PRB195205}
This correction was not applied to the -1 charged case, where the extra electron was delocalized and thus compensated the background charge without correction.
The stability of different charge states as a function of the Fermi level was calculated referencing the neutral charge state as zero of energy and the valence band maximum as zero of the Fermi level.

\begin{acknowledgement}

        The authors thank Churlhi Lyi and Youngkuk Kim for fruitful discussions.
        % K.M. and A.M.R. acknowledge support from the U.S. Department of Energy, Office of Science, Office of Basic Energy Sciences, under Award \#DE-SC0021118. Computational support was provided by the National Energy Research Scientific Computing Center (NERSC), a U.S. Department of Energy, Office of Science User Facility located at Lawrence Berkeley National Laboratory, operated under Contract No. DE-AC02-05CH11231.
        K.M. and A.M.R. acknowledge support from the U.S. Department of Energy, Office of Science, Office of Basic Energy Sciences, under Award \#DE-SC0026196. Computational support was provided by the National Energy Research Scientific Computing Center (NERSC), a U.S. Department of Energy, Office of Science User Facility located at Lawrence Berkeley National Laboratory, operated under Contract No. DE-AC02-05CH11231.
        K.M. acknowledges the JSPS Overseas Research Fellowship.

\end{acknowledgement}

\begin{suppinfo}

The following additional results are presented in the Supplementary Information.
\begin{itemize}
  \item Uniqueness of the \ch{Cs4Au3Cl12} structure
  \item Primitive cell structure and coordinates
  \item Full Born effective charges
  \item Spontaneous polarization branch
  \item COHP for all the nearest neighbors
  \item Comparison of the magnetic orders
  \item Polarization change induced by the \ch{Au^{2+}} current
  \item Effect of the \ch{Au^{2+}-Cl} bond length on \ch{Au^{2+}}
  \item Phonon dispersion
  \item Fermi energy dependence of \ch{Au^{2+}} stability
  % \item Topological analyses of the band structure
\end{itemize}
        Raw calculation inputs and outputs are available from (https://doi.org/10.5061/dryad.wwpzgmt0v).
\end{suppinfo}

\bibliography{reference}

\end{document}